\def\ln{\ell{n}}

\documentstyle[12pt]{article}

 {\parindent 0.5cm \textwidth 15.0cm \textheight
23.5cm \hoffset=-0.8cm \voffset=-3cm
  \let\LARGE=\large
 \let\large=\normalsize
\newcommand{\be}{\begin{equation}}
\newcommand{\ee}{\end{equation}}
\newcommand{\ba}{\begin{array}{c}}
\newcommand{\ea}{\end{array}}

\begin{document}
\begin{titlepage} \vspace{0.2in} \begin{flushright}
MITH-93/20 \\ \end{flushright} \vspace*{1.5cm}
\begin{center} {\LARGE \bf  Aspects of Dynamical Symmetry Breaking on a Lattice
\\} \vspace*{0.8cm}
{\bf She-Sheng Xue$^{a)}$}\\ \vspace*{1cm}
INFN - Section of Milan, Via Celoria 16, Milan, Italy\\ \vspace*{1.8cm}
{\bf   Abstract  \\ } \end{center} \indent

In the context of four-fermion and gauge interactions, the Schwinger-Dyson
equation for the fermion self-energy function is studied with the Wilson
fermion on a lattice. We find
that the critical line obtained by Bardeen, Leung and Love is
modified. No additional Goldstone bosons appear in
the spectrum of fundamental particles. Massive solutions to the
Schwinger-Dyson equation turn out to be possible for a small gauge coupling.
\vfill \begin{flushleft} 24th November, 1993   \vspace*{3cm} \\
\noindent{\rule[-.3cm]{5cm}{.02cm}} \\
\vspace*{0.2cm} \hspace*{0.5cm} ${}^{a)}$
E-mail address: xue@milano.infn.it\end{flushleft} \end{titlepage}

\noindent
{\bf 1.}\hspace*{0.3cm} Dynamical symmetry breaking plays an important role in
understanding success of the Standard Model. The structure of dynamical
symmetry breaking has been studied intensively for a wide variety of
theoretical models. Due to the lack of clear ideas as to the physical origin
of dynamical symmetry breaking in the Standard Model and the essential
difficulties in attempting to study such non-perturbative phenomena, progress
is still quite limited. One theoretical approach to the study of this
phenomenon is to find the solutions of the Schwinger-Dyson equation (SDE) for
the
fermion self-energy function $\Sigma(p)$. For a homogeneous SDE
without an explicit ultraviolet cutoff ($\Lambda$), it was shown \cite{John}
that the mass operator ($\bar\psi\psi$) has an anomalous dimension
$d_{\bar\psi\psi}=2+\sqrt{1-3\alpha/\pi}$ for a finite gauge coupling and thus
the high momentum decrease of solutions $\Sigma(p)\rightarrow
(p^2)^{-3+d_{\bar\psi\psi}}$ as $p^2\rightarrow\infty$. Further work
\cite{maskawa} in a version of the theory with an explicit ultraviolet cutoff
shows: (i) there are (no) spontaneous symmetry breaking solutions to the
homogeneous SDE for gauge coupling $\alpha > {\pi\over 3}$ ($\alpha <
{\pi\over 3}$); (ii) for an inhomogeneous SDE, all solutions for weak
coupling require that the inhomogeneous term, i.e. an explicit bare fermion
mass
$m_0$, goes as $m_0\sim \Lambda^{-(1-\sqrt{1-3\alpha/\pi})}$ in the
continuum limit ${\Lambda\over m}\gg 1$, where $m$ is an infrared
mass scale for the theory. However the mass operator
$m_\circ\bar\psi\psi$ remains finite at this limit and generates explicit
breaking of chiral symmetry.

The solutions to the SDE were analyzed in detail for both weak and
strong couplings \cite{kugo}. The critical point $\alpha_c=\pi/3$ is viewed
\cite{mira} as an ultraviolet fixed point of the theory in order to give an
sensible infrared limit for the theory. However, the renormalization properties
of the theory are called into question for the required running of the gauge
coupling with the cutoff. The most important progress made by Bardeen, Leung
and Love (BLL)\cite{bardeen} is to consider a system of gauge interaction and
four-fermion interaction, which is induced by integrating out the
high-frequence contribution of the theory. The theory was
shown to exhibit a line of critical points separating the chirally broken phase
from the symmetric one. The renormalization properties of the theory , the
relevance of the four-fermion operators and the spontaneous breaking of chiral
symmetry with accompanying pseudo-Goldstone bosons were intensively studied. In
this
paper, using the gauge-invariant Wilson lattice regularization \cite{wilson},
we
study the SDE for the fermion self-energy function in the system
containing gauge interaction and
four-fermion interactions. As an initial step, we show a modified BLL
critical line, where a sensible infrared limit of the theory can
probably be realized.

\vspace*{0.5cm}
\noindent
{\bf 2.}\hspace*{0.3cm} We are thus led to consider the following lagrangian
with the Wilson lattice regularization
\begin{eqnarray}
S&=&S_g(U)+{1\over2a}\sum_{n,\mu}\bar\psi_n\gamma^\mu(\psi_{n+\mu}-\psi_{n-\mu})+
\mu_0\sum_n\bar\psi_n\psi_n\nonumber\\
&&+{r\over a}\sum_{n,\mu}\bar\psi_n(U_{n,\mu}\psi_{n+\mu}
+U^\dagger_{n,\mu}\psi_{n-\mu}-2\psi_n)
+G\sum_n\big((\bar\psi_n\psi_n)^2+(\bar\psi_n\gamma_5\psi_n)^2\big),
\label{action}
\end{eqnarray}
where $S_g(U)$ is a pure gauge action ($U\in SU(N_c), U_{em}(1)$), $\mu_0$ is
the bare fermion mass, $a={\pi\over\Lambda}$ is the lattice spacing, and the
last term represents the Nambu-Jona Lasinio
fermion interactions \cite{nambu} with strength $G$. The fourth term is the
Wilson term with the Wilson parameter $r$, which cannot be avoided in the
lattice
regularization \cite{nogo}. The origin of this effective lagrangian
(\ref{action}), which may be induced by either the quantum gravity \cite{xue}
or high frequence contributions of the gauge theory, will not be a
focus of this paper.

Owing to the Wilson term and the bare fermion mass, chiral symmetry is
explicitly broken in (\ref{action}) and thus nothing can prevent the
generation the dimension-three mass operator $m\bar\psi\psi$. The
point is whether we can have a sensible infrared limit ${\Lambda\over m}\gg 1$
of the theory. In order to see how this happens, we study the SDE of
the fermion self-energy function $\Sigma(p)$ in the quenched and
planar approximations. As illustrated in Fig.1, we have
\begin{eqnarray}
\Sigma(p)&=&\mu_0+2g\int_q{\Sigma(q)+{r\over a}w(q)\over s^2(q)+M(q)^2}
+{\lambda\over a}\int_q
{1\over 4s^2({p-q\over 2})}\left(\delta_{\mu\nu}-\xi{s_\mu({p-q\over 2}))
s_\nu({p-q\over 2})\over
s^2({p-q\over 2})}\right)\nonumber\\
&&\cdot\Big(V^{(2)}_{\mu\nu}(p,p)-V^{(1)}_\mu(p,q){1\over \gamma_\rho s_\rho(q)
+M(q)}V_\nu^{(1)}(p,q)\Big),
\label{sd}
\end{eqnarray}
where $p (q)$ are dimensionless external (internal) momenta;
$s_\mu(l)=\sin(l_\mu)$ and $s^2(l)=\sum_\mu\sin(l_\mu)$; $\lambda=e^2
({e^2(N^2-1)\over 2N})$ for $U_{em}(1)(SU(N))$ gauge group and $ga^2=GN$; the
Wilson term $w(q)=\sum_\mu(1-\cos q_\mu)$ and $M(q)=a\Sigma(q)+rw(q)$;
$\int_q=\int^\pi_\pi {d^4q\over (2\pi)^4}$. The vertices \cite{smit} are
($k_\mu={(p_\mu+q_\mu)\over 2}$)
\begin{equation}
V_\mu^{(1)}(p,q)=\left(\gamma_\mu\cos k_\mu+
r\sin k_\mu\right);\hskip0.2cm
V_{\mu\nu}^{(2)}(p,q)=a\left(-\gamma_\mu\sin k_\mu+
r\cos k_\mu\right)\delta_{\mu\nu}.
\label{vertex}
\end{equation}
First of all, we need to analyze consistent solutions to the SDE
(\ref{sd}) on the lattice.

One of the main novelties of (\ref{sd}) is the non trivial interplay between
the continuum region, i.e. for momenta ($qa\ll 1$), and the truly discrete
region $qa\simeq 1$. In order to study such interplay it is important to
introduce a ``dividing scale'' $\epsilon$, such that $ma\ll\epsilon\ll \pi$.
Separating the integration region in (\ref{sd}) into two regions the
``continuum region'' $(0,\epsilon)^4$ and the ``lattice region''
$(\epsilon,\pi)^4$, we may separate our integral equations into the
``continuum part'' and the ``lattice part''
\begin{eqnarray}
\Sigma(p)&=&\mu_0+2G\int_{\epsilon\Lambda}{d^4q'\over (2\pi)^4}
{\Sigma(q')\over (q')^2+\Sigma(q')^2}
+2g\beta_1(r,\epsilon)+{r\over a}\beta_2(r,\epsilon)\nonumber\\
&&+{\alpha\over \alpha_c}\int_{\epsilon\Lambda}
{d^4q'\over 4\pi^2}{1\over
(p'-q')^2}{\Sigma(q')\over (q')^2+\Sigma(q')^2}
+{\alpha\over \alpha_c}\delta_1(r,\epsilon)+{r\over a}\delta_2(r,\epsilon),
\label{separation}
\end{eqnarray}
where $p',q'$ are dimensionful momenta; $\alpha={\lambda\over 4\pi},
\alpha_c={\pi\over 3}$ and
the ``continuum part''. Eq.(\ref{separation}), where the Landau gauge $\xi=1$
is chosen, is same as that derived from the continuum theory with an
intermediate cutoff $\epsilon\Lambda$. As
for the contributions to the integral equation from
the discrete ``lattice region''
$\beta_i(r,\epsilon),\delta_i(r,\epsilon) (i=1,2)$, we obtain:
\begin{eqnarray}
\beta_1(r,\epsilon)&=& \int_{q\in (\epsilon,\pi)^4}
{\Sigma(q)\over s^2(q)+M(q)^2}\label{b1}\\
\beta_2(r,\epsilon)&=& 2g\int_{q\in (\epsilon,\pi)^4}
{w(q)\over s^2(q)+M(q)^2}\label{b2}\\
\delta_1(r,\epsilon)&\simeq&
-\int_{(\epsilon,\pi)^4}{d^4q\over 4\pi^2}{\Sigma(q)\over
4s^2({q\over 2})}
\left[{-c^2({q\over 2})+r^2s^2({q\over 2})\over
s^2(q)+M^2(q)}\right]\label{b3}\\
\delta_2(r,\epsilon)&\simeq& {\alpha\over\alpha_c}\int_{q\in (\epsilon,\pi)^4}
{d^4q\over 12\pi^2}\!{1\over (4s^2({q\over 2}))}\left[{1\over 2}-
{w(q)(-c^2({q\over 2})+r^2s^2({q\over 2}))+s^2
(q)\over s^2(q)+M^2(q)}\right]\label{b4}
\end{eqnarray}
where $c^2(l)=\sum_\mu\cos^2(l_\mu)$.
The dependence on the
external
momentum $p\in (0,\epsilon)^4$ is omitted in $\delta_i(r,\epsilon),
$ because $p\ll q$ in the ``lattice region'' $q\in (\epsilon,\pi)^4$.
Note that (i) $\delta_i(r,\epsilon)$ do not depend on
the gauge parameter $\xi$ for there is a perfect cancellation
between the ``contact'' and the
``rainbow'' diagrams (Fig.1), which is guaranteed by Ward's identities;
(ii) the limits
$\lim_{\epsilon\rightarrow 0}\delta_2(r,\epsilon)$ and
$\lim_{\epsilon\rightarrow 0}\beta_2(r,\epsilon)$ can be taken since the
functions
$\delta_2(r,\epsilon)$ and $\beta_2(r,\epsilon)$ are regular in the limit
$\epsilon\rightarrow 0$, while this is not case for the functions
$\delta_1(r,\epsilon)$ and $\beta_1(r,\epsilon)$; (iii) within the ``lattice
region'',
the Wilson parameter $r$ must not vanish and all functions
$\beta_i(r,\epsilon)$
and $\delta_i(r,\epsilon)$ remain non-vanishing.

\vspace*{0.5cm}
\noindent
{\bf 3.}\hspace*{0.3cm} The terms $\delta_2(r,0)$, $\beta_2(r,0)$ and
the bare fermion mass
$\mu_0$ of the SDE (\ref{sd}) diverge as ${1\over a}$. For the purpose
of finding the ``continuum'' counterpart of the theory where $\Sigma(p)a\ll 1$,
we make a fine-tunning of the bare fermion mass $\mu_0$ such that
\begin{eqnarray}
\Sigma(p)&=&2G\int_{\epsilon\Lambda}{d^4q'\over (2\pi)^4}
{\Sigma(q')\over (q')^2+\Sigma(q')^2}
+2g\beta_1(r,\epsilon)\nonumber\\
&&+{\alpha\over \alpha_c}\int_{\epsilon\Lambda}
{d^4q'\over 4\pi^2}{1\over
(p'-q')^2}{\Sigma(q')\over (q')^2+\Sigma(q')^2}
+{\alpha\over \alpha_c}\delta_1(r,\epsilon)
\label{continuum}\\
0&=&\mu_0+{r\over a}\beta_2(r,0)+{r\over a}\delta_2(r,0).
\label{fine}
\end{eqnarray}
This fine-tuning (\ref{fine}), which can also be called the ``chiral limit'',
can probably be
guaranteed \cite{rome} by chiral symmetry at short distances and we shall
not discuss it in this paper. It is self-consistent that once the
renormalization
(\ref{fine}) is performed, the fermion self-energy function $\Sigma(p)$
contained in
the numerators and denominators of eqs.(\ref{b1},\ref{b2},\ref{b3},\ref{b4})
and (\ref{continuum})
is free from ${1\over a}$ divergence
$a\Sigma(q)\simeq 0, M(q)^2\simeq (rw(q))^2$. We thus obtain the ``continuum''
counterpart of the SDE (\ref{continuum}).

Let us now address the important question of the $\epsilon$-independence of our
results. The introduction of the ``dividing scale'' in (\ref{sd}) is,
apart from
the requirement $ma\ll\epsilon\ll\pi$, rather arbitrary, thus no dependence
on $\epsilon$ should appear in our final results.
In order for such independence to occur, as it must, it
is clear that the $\epsilon$-dependent terms (e.g., $\ln\epsilon$) from the
continuum integral in
(\ref{continuum}), must be compensated by analogous
terms arising in the calculation of $\delta_1(r,\epsilon)$ and
$\beta_1(r,\epsilon)$. Owing to integral momenta $q\in (\epsilon,\pi)^4$ in
eqs.(\ref{b2}),(\ref{b4}), we can make the reasonable
approximation $\Sigma(q)\simeq\Sigma
(\Lambda)$ in the numerators of $\delta_1(r,\epsilon)$ and
$\beta_1(r,\epsilon)$. Thus the $\epsilon$-independent terms
$\Sigma(\Lambda)\delta_0(r) (\Sigma(\Lambda)\beta_0(r))$ contained in
$\delta_1(r,\epsilon) (\beta_1(r,\epsilon))$ can be found by
numerical calculation. The numerical functions $\delta_0(r)$ and $\beta_0(r)$
are reported in Fig.2 and 3. Thus segregating
the $\epsilon$-dependent terms in (\ref{continuum}), we may write
\begin{eqnarray}
\Sigma(p)&=&2G\int_{\Lambda}{d^4q'\over (2\pi)^4}
{\Sigma(q')\over (q')^2+\Sigma(q')^2}
+2g\Sigma(\Lambda)\beta_0(r)\nonumber\\
&&+{\alpha\over \alpha_c}\int_{\Lambda}
{d^4q'\over 4\pi^2}{1\over
(p'-q')^2}{\Sigma(q')\over (q')^2+\Sigma(q')^2}
+{\alpha\over \alpha_c}\Sigma(\Lambda)\beta_0(r),
\label{fingap}
\end{eqnarray}
which is analogous to the ``chiral limit'' SDE of the continuum theory
with two additional boundary terms $2g\Sigma(\Lambda)\beta_0(r)$ and
${\alpha\over \alpha_c}\Sigma(\Lambda)\delta_0(r)$.

\vspace*{0.5cm}
\noindent
{\bf 4.}\hspace*{0.3cm} It is straightforward to adopt the analysis of
Bardeen, Leung and
Love \cite{bardeen}. After performing the angular integration and changing
variables to $x=(p')^2$, eq.(\ref{fingap}) becomes a boundary-value problem,
\begin{eqnarray}
{d\over dx}(x^2\Sigma '(x))+{\alpha\over 4\alpha_c}{x\over x+\Sigma^2(x)}
\Sigma(x)&=&0\nonumber\\
(1+\tilde g)\Lambda^2\Sigma '(\Lambda)+\Sigma(\Lambda)(1-2\tilde g
{\alpha\over\alpha_c}\beta_0(r)
-{\alpha\over \alpha_c}\delta_0(r))&=&0,
\label{diff}
\end{eqnarray}
where $\tilde g=Ga^{-2}N{\alpha_c\over\alpha}$. The solution to this boundary
value problem is well established. For weak coupling, we
have the gap equation,
\begin{eqnarray}
\tanh\theta &=& {(\tilde g+1)\sqrt{1-{\alpha\over\alpha_c}}\over
(\tilde g-1)+4\tilde g{\alpha\over\alpha_c}\beta_0(r)+2{\alpha\over\alpha_c}
\delta_0(r)}\\
\theta &=& \sqrt{1-{\alpha\over\alpha_c}}\ln\big({\Lambda \over m}\big),
\label{gap}
\end{eqnarray}
where $m=\Sigma(0)$. There is a critical line along which a sensible infrared
limit of the theory can be defined. In this limit, ${\Lambda\over m}\gg 1$
and $\theta\gg 1$, so that the critical line relates $\tilde g$ and $\alpha$ as
\begin{equation}
\tilde g = {(1+\sqrt{1-{\alpha\over\alpha_c}})-
2{\alpha\over\alpha_c}\delta_0(r)
\over (1-\sqrt{1-{\alpha\over\alpha_c}})+4{\alpha\over\alpha_c}\beta_0(r)},
\label{critical}
\end{equation}
which is reported in Fig.~4 with $r=0.2, 0.6, 1.0$. We find that (i) the BLL
critical
line is modified by the ``lattice'' terms $\beta_0(r)$ and $\delta_0(r)$ with
$0< r\leq 1$; (ii) $\tilde g={1-2\delta_0(r)\over 1+4\beta_0(r)}$ for $\alpha
=\alpha_c$ (the MBLL critical point was $\tilde g=1$ for $\alpha =\alpha_c$);
(iii) $\alpha =0$ and $\tilde g{\alpha\over\alpha_c} ={4\over 1+8\beta_0(r)}$
(the NJL critical point).
For $\tilde g = 0$, we need to have $2{\alpha \over\alpha_c}\delta_0(r)> 1,
\delta_0(r)> 0$ (see eq.(\ref{gap})), and the critical line (\ref{critical})
gives us a critical point at ${\alpha \over\alpha_c}={4\delta_0(r)-1\over
(2\delta_0(r))^2}$. This leads to $2\delta_0(r)> 1$, for which there is no room
for the values of $r$ (see Fig.2).
We also find there is no room
($\tilde g =0$) for the existence of a very small critical value of the gauge
coupling ${\alpha \over\alpha_c}\ll 1$, unless $\delta_0(r)\gg 1$.

\vspace*{0.5cm}
\noindent
{\bf 5.}\hspace*{0.3cm} We observe that the contributions from the ``lattice''
region, where the Wilson term survives, have an impact on the SDE for the
fermion self-energy function, thus the BLL critical line obtained from the
continuum limit is modified. The modification of the various critical exponents
and critical scaling laws will be discussed in a future work. We show that it
is worthwhile to study the aspects of dynamical symmetry breaking on the BLL
approach with gauge-invariant Wilson lattice regularization.

The functions $\delta_0(r)$ and $\beta_0(r)$ are calculated in the
planar approximation and they are expected to change if we take vertex
contributions into account. The quenched approximation is also adopted in this
work, $\alpha$ is expected to be renormalized with the renormalization constant
$Z_3$ if we consider the fermion-loop contributions.

There are no pseudo-Goldstone bosons appearing in the
spectrum of the theory since chiral symmetry is explicitly broken in
lagrangian (\ref{action}). The disappearance of extra Goldstone bosons is
significant phenomenologically in the Standard Model.
The ``chiral limit''
(\ref{fine}) means that the bare fermion mass term $\mu_0\bar\psi\psi$ is
treated
as a mass counterterm to eliminate ${1\over a}$ divergences stemming from the
Wilson term so that we can have the continuum limit ${\Lambda\over m}\gg 1$ of
the theory. Future work is necessary to discuss how these breaking terms
(the bare fermion mass term and the Wilson
term) and the effective four-fermion interactions are generated for a theory
possessing perfectly chiral symmetry at short distances.

\newpage  \pagestyle{empty}
\begin{center} \section*{Figure Captions} \end{center}

\vspace*{1cm}

\noindent {\bf Figure 1}: \hspace*{0.5cm}
 The quenched and planar approximated Schwinger-Dyson equation.

\noindent {\bf Figure 2}: \hspace*{0.5cm}
 The function $\beta_0(r)$ in terms of the Wilson parameter $r$.

\noindent {\bf Figure 3}: \hspace*{0.5cm}
 The function $\delta_0(r)$ in terms of the Wilson parameter $r$.

\noindent {\bf Figure 4}: \hspace*{0.5cm}
 The critical line (\ref{critical}) for $r=0.2, 0.6, 1.0$ ($g_c\equiv
\tilde g{\alpha\over\alpha_c})$.

}


\begin{thebibliography}{99}

\bibitem{John}
K.~Johnson, M.~Baker and R.~Willey, {\sl Phys.~Rev.} {\bf 136}
(1964) 1111, {\it ibid.} {\bf 163} (1967) 1699;\\
S.L.~Adler and W.A.~Bardeen, {\sl Phys.~Rev.} {\bf D4} (1971) 3045.

\bibitem{maskawa}
T.~Maskawa and H.~Nakajima, {\sl Prog.~Theo.~Phys.} {\bf 52} (1974) 1326 and
{\it ibid.} {\bf 54} (1975) 860.

\bibitem{kugo}
R.~Fukuda and T.~Kugo, {\sl Nucl.~Phys.} {\bf B117} (1976) 250;\\
K.~Higashijima and A.~Nishimura, {\sl Nucl.~Phys.} {\bf B113} (1976) 173;\\
T.~Akiba and T.~Yanagida, {\sl Phys.~Lett.} {\bf B169} (1986) 432.

\bibitem{mira}
V.A.~Miransky and P.I.~Fomin, {\sl Sov.~J.~Part.~Nucl.} {\bf 16} (1985) 203;
P.I.~Fomin, V.~Gusynin, V.A.~Miransky and Yu.~A.~Sitenko,
{\sl Riv.~Nuovo.~Cim.} {\bf 6} (1983)1.

\bibitem{bardeen}
W.A.~Bardeen, C.T.~Hill and M.~Linder,
{\sl Phys. Rev. Lett.} {\bf 12} (1986) 1230,
{\sl Nucl. Phys.} {\bf B233} (1986) 647 and {\it ibid.} {\bf B323} (1989) 493,
{\sl Phys. Rev.} {\bf D42} (1990) 3547;\\
A.~Koci\'c, S.~Hands, J.~Kogut and E.~Dagotto, {\sl Nucl.~Phys.} {\bf B347}
(1990) 217.

\bibitem {wilson}
K.~Wilson, in {\it New phenomena in subnuclear physics\/}
(Erice, 1975)
ed.\ A.~Zichichi (Plenum, New York, 1977).

\bibitem{nambu}
Y.~Nambu and G.~Jona-Lasinio, {\sl Phys. Rev.} {\bf 122} (1961) 345.

\bibitem{nogo}
H.B.~Nielsen and M.~Ninomiya, {\sl Nucl.~Phys.} {\bf B185} (1981) 20, {\it
ibid.} {\bf B193} (1981) 173, {\sl Phys.~Lett.} {\bf B105} (1981) 219.

\bibitem{xue}
G.~Preparata and S.-S.~Xue, {\sl Phys.~Lett.} {\bf B264} (1991) 35,
{\it ibid.} {\bf B302} (1993) 443,
{\sl Nucl.~Phys.} {\bf B30} (Proc.~Suppl.) (1993) 647.

\bibitem{smit} L. H. Karsten and J. Smit, {\sl Nucl. Phys.} {\bf B144}
(1978) 536

\bibitem{rome}
A.~Borrelli, L.~Maiani, G.C.~Rossi, R.~Sisto and M. Testa, {\sl Nucl.~ Phys.}
{\bf B333} (1990) 335, {\sl Phys.~Lett.} {\bf B221} (1989) 360.

\end{thebibliography}
\end{document}